\documentclass[10pt,twoside]{article}

\usepackage{asp2006}
\usepackage{epsf}

\newcommand\bestfit{{\it best-fit}}

\newcommand\galphat{{\scshape Galphat}}

\newcommand\bt{{\it B/T}}
\newcommand\sn{{\it S/N}}

\newcommand\Ks{$K_s$}

\newcommand\etal{et al.}

\newcommand\fig{{\scshape Figure}}

\newcommand\tab{{\scshape Table}}

\newcommand\haussler{H\"{a}u{\ss}ler}
\newcommand\sersic{S\'{e}rsic}

\begin{document}

\title{Beyond the {\it best-fit} parameter: new insight on galaxy
  structure decomposition from GALPHAT}

\author{Ilsang Yoon, Martin D. Weinberg and Neal S. Katz}

\affil{Department of Astronomy, University of Massachusetts, Amherst, USA}

\begin{abstract}
  We introduce a novel image decomposition package, \galphat, that
  provides robust estimates of galaxy surface brightness profiles
  using Bayesian Markov Chain Monte Carlo.  The \galphat-determined
  posterior distribution of parameters enables us to assign rigorous
  statistical confidence intervals to maximum a posteriori estimates
  and to test complex galaxy formation and evolution hypotheses.  We
  describe the \galphat~algorithm, assess its performance using test image
  data, and demonstrate that it has sufficient speed for production analysis of a
  large galaxy sample. Finally we briefly introduce our ongoing
  science program to study the distribution of galaxy structural
  properties in the local universe using \galphat.
\end{abstract}

\section{What and Why is  GALPHAT?}

Large photometric and spectroscopic surveys of galaxies
[e.g. SDSS\cite{sdss} and 2MASS\cite{tmass}] continue to provide vast
ensembles of galaxy images, but rigorous conclusions about galaxy
formation and evolution hypotheses based on the full volume of
information present serious computational and algorithmic challenges.
For example, parametric models are widely used to derive galaxy
structural parameters: brightness, size, profile shape and
ellipticity.  A recent study \citep{allen} presented a bulge-disc
decomposition for $10^4$ nearby galaxies.  However, an accurate
decomposition is stymied by degeneracies in the parameter estimation
itself.  A \sersic~profile has 8 parameters, and the topology of the
likelihood function in this high-dimensional parameter space is 
too complex to visualize and hard to characterize robustly.  In most 
previous galaxy decomposition analyses, the uncertainties of
derived model parameters have not been carefully characterized.  The
correlations of physical properties and structural parameters of
galaxies are usually assessed through {\it scatter} plots of the
\bestfit~parameters (e.g. maximum a posteriori estimates with formal
inverse Hessian variance estimates). These correlations are subject to
strong contamination by underlying systematic correlations of each
model parameter.

A Bayesian approach to parameter inference and hypothesis testing
naturally addresses these difficulties.  The probability of model
parameters ($M$) for a given data set ($D$), $P(M|D)$, is related to
the probability of the data given the model (the likelihood),
$P(D|M)$, and prior probability of the model, $P(M)$, through Bayes
theorem:
\begin{equation}
  P(M|D) = \frac{P(D|M) P(M)}{P(D)} = \frac{P(D|M) P(M)}{\int P(D|M) P(M) dM}
\label{bayes}
\end{equation}
In our galaxy decomposition problem, $M$ is the vector of parameters
describing the model and $D$ is the image data.  For a
high-dimensional model space, direct integration is infeasible, and
one resorts to Markov Chain Monte Carlo (MCMC) methods.  Although
costly, sampling the posterior using MCMC returns a robust estimate of
the model parameters.  From this distribution, one may also define a
\bestfit~parameter value and confidence bounds, but the real power in
this approach is the natural description by the posterior distribution of
any intrinsic correlation between model parameters.  This power comes
at a cost: Bayesian MCMC requires much more computation time than
$\chi^2$ minimization.  To make this feasible for present-day surveys
with large numbers of galaxies, we present a novel image decomposition
package \galphat~(GALaxy PHotometric ATtributes
\footnote{\galphat~web page: http://sites.google.com/site/galphat/galphat}) which uses a
state-of-the-art Bayesian computation software package BIE (Bayesian
Inference Engine\footnote{For more information on BIE, see the BIE web
  page\,\citep{bieweb} and \cite{weinberg}}) for sampling MCMC
efficiently and incorporates an optimized image processing algorithm
to reduce the likelihood evaluation time.  We will introduce
the algorithm, test its performance, and present science using
\galphat~in the following sections. 

\section{The GALPHAT Algorithm}

Given a parameter vector $M$, \galphat~produces a likelihood function
for an image (pixel data, mask, PSF, background) and the BIE samples
the posterior for a prior distribution using an MCMC algorithm.  The
BIE provides a choice of MCMC algorithm depending on the complexity of
the problem.  To date, we have found that a multiple chain algorithm
with tempering \citep{geyer} is a good compromise between speed and
flexibility.  For the tests described here, \galphat~models the galaxy surface
brightness using multiple 2D \sersic~profiles with arbitrary
ellipticities and position angles.  For computational efficiency,
\galphat~pre-generates two-dimensional cumulative distributions of
\sersic~profile with many different $n$ using a rigorous error
tolerance. By assigning pixel values by table interpolation,
\galphat~generates a model image with arbitrary axis ratio $b/a$.
Image rotation is done in Fourier space using three shear operations;
a rotation matrix is decomposed into three sequential shear matrices
in the X,Y, and X directions and then each shear operation is carried
out using the Fourier shift theorem \citep{larkin}.  Then, the model
image is convolved with a given PSF and an adjustable flux pedestal
with a spatial gradient is added to model the sky background.
\galphat~uses a Poisson likelihood function to describe the photon
counting process.  A more detail description of the algorithm will be
published in upcoming method papers \citep{galphat} and
\citep{weinberg}.

\section{GALPHAT performance}
For testing, we generate an ensemble of synthetic galaxy images over
a wide range of signal-to-noise ratios \sn~and sizes using the IDL program by
\haussler\citep{haussler}. \galphat~successfully recovers the input
parameters with robust statistical confidence intervals (CL).

\begin{figure}
\input{epsf}
\epsfxsize=3truein
\centerline{\epsfbox{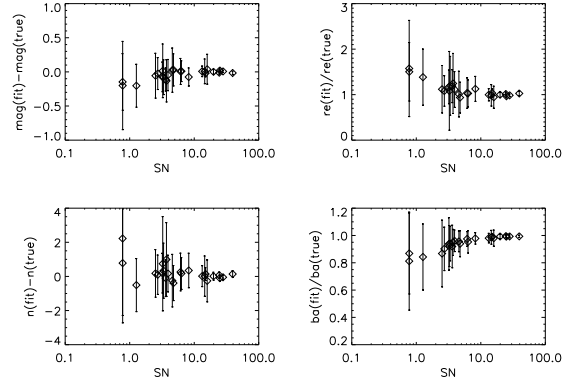}}
\vspace{3mm}
\caption{The differences between \galphat~posterior medians and
  the exact values as a function of \sn~for \sersic~$n=4$ profiles with
  $b/a=1$.  Error bars are 99.7\% CL.  The \galphat~inference is
  unbiased until \sn$\approx3$.  The systematically lower estimate of $b/a$
  owes to the uniform prior with a range of [0,1].}
\label{snpar}
\end{figure}

\fig~\ref{snpar} compares the \galphat~posterior median and 99.7\% CL
with the true magnitude, effective radius $r_e$, \sersic~index $n$ and
axis ratio $b/a$ as a function of \sn.  We define \sn~as the ratio of
the average flux per pixel from the source within $r_e$ to the
average flux from noise [following
\haussler~\etal\citep{haussler}]. We use 10,000 converged MCMC samples
for characterizing the posterior.  As \sn~decreases, the 99.7\% CL
range increases.  However \galphat~is robust and encloses true
parameter within the 99.7\% CL even in the extreme case of \sn$\sim 1$.

\begin{figure}
  \input{epsf} \epsfxsize=4truein \centerline{\epsfbox{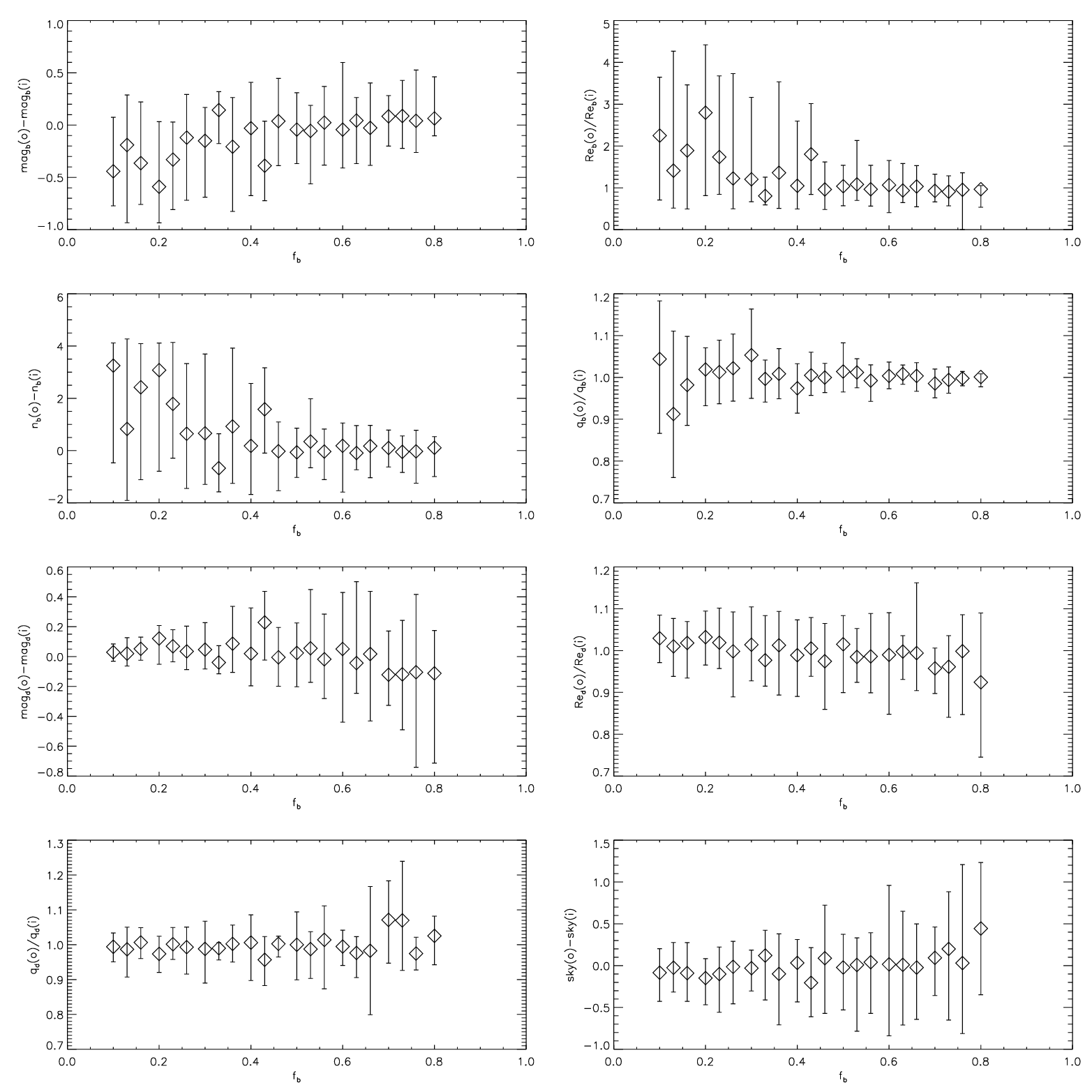}}
  \vspace{3mm}
  \caption{The differences between \galphat~posterior medians and
    the exact values for bulge-disk galaxies as a function of
    $f_b=$\bt. Error bars are 99.7\% CL. From the top left to the
    bottom right panel, corresponding parameters are bulge magnitude,
    bulge $r_e$, bulge $n$, bulge $b/a$, disk magnitude, disk $r_e$,
    disk $b/a$, and sky background.  }
\label{fbpar}
\end{figure}

To assess the performance for bulge-disk decomposition, we use
 simulated galaxies with a \sersic~bulge ($n=4$) and an
exponential disk ($n=1$).  \fig~\ref{fbpar} compares the
\galphat~parameter inferences to the exact values with varying
bulge-to-total flux ratio \bt.  As the bulge fraction increases, the
inference of bulge parameters becomes more reliable and the inference of
disk parameters becomes less reliable, as expected.  The marginalized
posterior distribution for pairs of parameters show that correlations
between parameters abound; the magnitude of the correlations depend on
\bt.  Clearly, a characterization of this parameter correlation is
necessary for interpreting the distribution of these parameters for
the galaxy population overall and for testing hypotheses of galaxy
formation and evolution.  \galphat~provides the full posterior
probability distribution of model parameter and enables us to quote
reliable uncertainties for each model parameter by including all the
systematic correlations.  Furthermore, the posterior distribution enables
the computation of higher-order statistics and more general model
comparisons.
  
Convergence may be slow in high-dimensional parameter spaces owing to
the complexity of the likelihood function.  We have found it
productive to iteratively add image information using a hierarchy of
successively aggregated images.  Beginning with the most aggregated
image (Level 0) one computes the posterior, $P(\theta|D_0)$.  The
posterior for the next level (Level 1) is $P(\theta|D_1) \propto
P(\theta|D_0) [P(D_1|\theta)/P(D_0|\theta)]$ and so on.  Using this
hierarchical data structure, \galphat~reduces the run time by factors
of two, depending on the level of aggregation, by accelerating
convergence.  Also, the BIE checkpoints the full state of the
simulation, efficiently enabling the posterior distribution to be
augmented at a later date.

\galphat~can be run on either a single CPU or in a cluster environment.
\tab~\ref{tab} provides \galphat~runtimes for a simple one-component
decomposition.  Of course, in addition to computing hardware, the time
for your parameter estimate will depend on image size, characteristics
of the model, the MCMC algorithm, the convergence diagnostic method, and
the required number of posterior samples. 
More detailed results will be shown in the methods paper\citep{galphat}. 
\begin{table}
\centering
\center
\caption{\scshape{wall clock time for \galphat}}
\centering
\vspace{1truemm}
\begin{tabular}{@{}llllll@{}} \hline\hline
Image & samples & CPU & Processors & Level &runtime \\ \hline
200 by 200     & 20,000  & AMD Athlon(tm) MP 1800+ & 8  & 1 & 2 hr \\
single \sersic &         & 1533 MHz                &    &  &         \\ \hline
200 by 200     & 20,000  & AMD Athlon(tm) MP 1800+ & 8  & 2 & 40 min \\
single \sersic &         & 1533 MHz                &    &  &        \\ \hline
\end{tabular}
\label{tab}
\end{table}

\section{Ongoing science program using GALPHAT}
We are currently performing a bulge+disk model decomposition of 2MASS \Ks
band selected galaxies with $7<K_s<11$ to derive luminosity functions
for each component in the present day Universe.  We will investigate
the effect of environment on these distributions.  From these, we can
compare to mass assembly histories predicted from hierarchical galaxy
formation theories.

\acknowledgements We thank Dr. Daniel McIntosh and Mr. Yicheng Guo for
useful discussions and providing sample images in early stage of this
work.  This work has been supported in part by the NSF Grant IIS
0611948 and by the NASA AISR Program Grant NNG06GF25G.


\begin{thebibliography}{}
\bibitem[1]{allen} 1.\quad P. Allen~\etal, 2006, MNRAS, 371, 2 
\bibitem[2]{haussler}2.\quad B. \haussler~\etal, 2007, ApJS, 172, 615
\bibitem[3]{larkin}3.\quad K. G. Larkin, M. A. Oldfield, H. Klemm, 1997, Optics Communications 
\bibitem[4]{tmass}4.\quad M. Skrutskie~\etal, 2006, AJ, 131, 1163
\bibitem[5]{geyer}5.\quad C. J. Geyer, E. A. Thompson, 2005, JASA, 90, 909
\bibitem[6]{galphat}6.\quad I. Yoon, M. D. Weinberg, N. S. Katz, 2009, in preparation
\bibitem[7]{sdss}7.\quad D. York~\etal, 2000, AJ, 120, 1579
\bibitem[8]{bieweb}8.\quad http://www.astro.umass.edu/BIE
\bibitem[9]{weinberg}9.\quad M. D. Weinberg~\etal, 2009,  in preparation
\end{thebibliography}
\end{document}